\documentclass[nofootinbib,prd,amssymb,12pt]{revtex4}

\newcommand{\beq}[1] {\begin{equation}\label{#1} }
\newcommand{\eeq} {\end{equation} }

\newcommand{\bea}[1]{\begin{eqnarray}\label{#1} }
\newcommand{\eea}{\end{eqnarray}}

\newcommand{\al}{\alpha}
\newcommand{\be}{\beta}

\newcommand{\del}{\delta}
\newcommand{\Del}{\Delta}
\newcommand{\lam}{\lambda}
\newcommand{\sig}{\sigma}

\newcommand{\der}{\partial}
\newcommand{\Dir}{\not{\partial}}
\newcommand{\dd}{\hbox{d}}

\begin{document}

\title{The Higgs sector on a two-sheeted space time}
\author{Cosmin Macesanu\footnote{cmacesan@physics.syr.edu} and 
Kameshwar C. Wali \footnote{wali@physics.syr.edu}}
\affiliation{Department of Physics \\
Syracuse University \\
Syracuse, NY 13244-1130, USA}

\begin{abstract}
We present a general formalism based on the 
framework of non-commutative geometry, suitable to the study the standard model 
of electroweak interactions, as well as that of more general
gauge theories. Left- and right-handed 
chiral fields are assigned to two different sheets of 
space-time (a discretized version of Kaluza-Klein theory). 
Scalar Higgs fields
find themselves treated on the same footing as the gauge fields, resulting in
spontaneous symmetry breaking in a natural and  predictable way. 
We first apply the formalism to the Standard Model, where one can predict
the Higgs mass and the top Yukawa coupling. 
We then study the left-right symmetric model, where we show that this
framework imposes  constraints on the type and coefficients
of terms appearing in the Higgs potential.


\end{abstract}

\vspace*{-0.5cm}
\begin{flushright}
SU-4252-809
\end{flushright}
\vspace{0.5cm}

\maketitle

\section{Introduction}

The current picture of a continuum space-time at all scales
underlying a smooth manifold has proved inadequate to describe all
elementary particle interactions including gravity. The
mathematical frameworks of both general relativity and
quantum field theory of elementary particle interactions assume
such a smooth manifold, and their incompatibility is one of the
main signals for the inadequacy of such a continuum picture of
space-time. In search of a more general framework, Connes has
proposed an alternate approach based on 
non-commutative geometry (NCG) \cite{Connes,Connes2}. The basic idea of Connes
is to do away with the precise specification of an underlying
manifold as the starting point. Instead, he formulates its
description in terms of an associative and involutive algebra,
commutative or non-commutative. One may think of this as a
generalization of the well known theorem due to Gelfand, 
which states that the classical topological space based
on a continuum can be completely recovered from the Abelian
algebra of smooth functions.

Connes' ideas have been explored in several directions. Of
particular relevance is their application to the standard model
and beyond. In spite of the great success of the standard model in
confronting experiments, it is recognized that it is far from a
fundamental theory. One of the main drawbacks is that the mechanism of
spontaneous symmetry breaking depends on the {\it ad-hoc}
addition of a Higgs sector, with a somewhat arbitrary Higgs field content 
and an arbitrarily chosen potential. 
In contrast, Connes' approach has
given rise to a geometrical description of a gauge field theory in
which the Higgs fields finds themselves on the same footing as the
gauge fields, and where the Higgs potential has spontaneous symmetry
breaking built in it. 

In recent years, there has been a great deal
of work beginning with Connes and Lott 
(\cite{ConnesLott,Coquereaux,Chamseddine,Wulkenhaar} are some examples). 
However, the approaches used so far have tended to be highly
mathematical. Our approach will be somewhat heuristic and not
overly concerned with  mathematical details. We intend to keep
close to the familiar ideas of quantum field theories and
Riemannian geometry and use Connes' rigorous non-commutative
geometry as a guide to construct models with interesting
features from the physical point of view.

In our approach, 
the underlying manifold on which the theory is defined consists
of the direct product between the four-dimensional (4D)
Minkowski space-time and a 
finite number of points. Note that this is only a particular
(and rather simple) realization 
of the more general theory of Connes. Specifically, we consider
the case when the dimension of this discrete space is two; hence
the name two-sheeted space time. This framework allows a simple
physical interpretation, as a model in which 4D fields
live on two distinct branes embedded in a higher dimensional space;
in other words, a discretized version of Kaluza-Klein theory 
\cite{Viet,VietWali}.
As such, one can think that this theory might be derived as the
4D limit of a more fundamental theory (perhaps a string one).

We envisage the fermion left- and right-chiral fields living
 on
two separate sheets of space time
associated with the two discrete points.
The generalized gauge potential $\mathbb{A}$ (known as a connection in 
geometrical terms) appears as a two-by-two matrix
operator on this internal space.
The diagonal components couple
fermion fields living on the same sheet (therefore with the same
chirality) and consequently are identified with the usual 4D gauge fields.
The off-diagonal components couple left with right-handed fermion fields,
and therefore can be identified as standard Higgs fields. We show that it is
possible to build a suitable curvature  operator 
(the field strength $\mathbb{F}$),
and construct a consistent gauge theory by using these extended
gauge operators. The Higgs potential terms appear as intrinsic components
of the gauge invariant action for the Yang-Mills fields  
$*\mathbb{F}\mathbb{F}$.

We apply these ideas by first formulating the Standard Model as a 
$SU_C(3)\times SU_L(2)\times U_Y(1)$ gauge theory on two sheeted space-time.
Due to the simplicity of the Higgs sector in this case (one Higgs field
suffices), one finds that it is possible to predict the top Yukawa
coupling ($\sim$ 0.8, close to but not in exact agreement with the SM value),
as well as the Higgs mass. However,
we think that more interesting possibilities lie in the
application of this formalism  to theories which
may be valid at higher energy scales, such as grand unified theories
based on higher gauge groups. In such theories, generally the Higgs
sector can be quite complex, and in the absence of a guiding principle
one must rely on additional assumptions. 
 The approach described in this paper
  can provide such a guide.
 As an example, we formulate the 
left-right symmetric model in this framework.

The paper is organized as follows.
In Section II, we briefly review Connes' abstract algebra approach and
the concept of a spectral triple. In Section III, we adapt
this formalism to a two-sheeted space-time that may be considered
as a discretized version of Kaluza-Klein theory, in which the
compact circle in the fifth dimension is replaced by two discrete
points. 
Section IV and V are devoted to the construction of standard model and
the left-right symmetric model in this framework. The final
Section is devoted to a discussion of the results and conclusions.

\section{A brief review of non-commutative differential geometry.}

Connes'starting point of NCG is a universal
differential algebra $\Omega^{*} (\cal {A})$ constructed from an
associative, commuting or non-commuting algebra $\cal {A}$. It is
unital and involutive. This differential algebra can be thought of
as being generated by the elements $a,b$ and a symbol $\del$ with
properties $\del 1 = 0, \del(a b) = (\del a)b + a \del b$ for all
$a,b \in \cal {A}$. The zero order forms are simply the elements of
$\cal {A}$. Higher order forms in $\Omega^{p}\cal{A}$ are generated
by `extending' the differential $\del$ to an operator:
 if $a_0 \del a_1 \ldots \del a_p$ is a
$p-$form, then
$$ \del (a_0 \del a_1 \ldots \del a_p)  \ = \
\del a_0 \del a_1 \ldots \del a_p
$$
is a $(p+1)-$form. This also implies $\del^2 = 0$.
For a more detailed discussion at the introductory level, see, for example,
\cite{Schucker1}.

In physical applications of interest to us, the abstract algebra
becomes an algebra of operators acting on a Hilbert space
$\cal{H}$. The abstract elements $a\in\cal{A}$ are represented as
operators through a representation $\rho(a)\equiv A$ acting on
$\cal{H}$. The abstract symbol $\del$ becomes the exterior
derivative represented by a self-adjoint operator called the Dirac
operator $\cal{D}$. The three elements, the algebra $A$, the Dirac
operator $\cal{D}$ and the Hilbert space $\cal{H}$ together form a
\emph{spectral triple} according to Connes.

With the help of the Dirac operator, we can now extend the
representation of $\cal{A}$ to its algebra of differential forms
through the correspondence:
$$ \rho(a_0 \del a_1 \ldots \del a_p)  \ = \
\rho(a_0) [{\cal D},\rho(a_1)]  \ldots [{\cal D},\rho(a_p)] \ .$$

In this sense, the Dirac operator can be used to build a
`representation' of the formal exterior derivative $\del$ in the
space of operators acting on  ${\cal H}$; if $\phi \in
\Omega({\cal A})$:
$$ \rho(\del \phi ) \ = \ [{\cal D},\rho(\phi)]\ .$$
(Depending on the rank of the form $\phi$ and the grading properties of
the algebra,
one may also need to use anticommutators.)

One must be careful, however; if $\del$ is a proper exterior
derivative, one needs to have $\del (\del \phi) = 0$; but
generally $[{\cal D}, [{\cal D},\rho(\phi)]]$ is not zero. To get
around this problem, one defines an equivalence relation on the
spaces $\rho(\Omega^p({\cal A}))$, so that all the operators of
the form  $[{\cal D}, [{\cal D},\rho(\phi)]]$ are equivalent to
the zero operator; technically, the space of operators
corresponding to $p-$forms is $\rho(\Omega^p({\cal A}))$ divided
by the space of so-called junk forms ${\cal J} =
\rho(\del(\hbox{ker}\Omega^{p-1} ({\cal A}))$ (this is sometimes
called dividing out the junk)\footnote{A slightly different
definition of junk forms is used in the literature. However, for
our purposes, the two definitions seem to give the same results 
(see \cite{Schucker1}).}. 

 In our studies, $\cal{H}$ will consist of square integrable sections
of a spinor bundle representing physical states of the fermions.
Let us assume that the fermion fields form a basis for a
representation of some group $G$. For theories defined on a smooth
manifold, the gauge fields will then be associated with
differential one-forms defined on the manifold. The field
strength, which is used in the construction of the Lagrangian,
will be a two-form. Generally, physically relevant quantum
operators built out of gauge fields are elements of the
differential algebra $\Omega (M)$ defined on the manifold.

For the purposes of illustration, consider the simple case of
a gauge theory on the Minkowski manifold. The relevant algebra is
$C^{\infty}(M)$. The Hilbert space ${\cal H}$ is the space of
Dirac spinors; and the Dirac operator is the standard ${\cal D} =
\not{ \partial} = \gamma^\mu \der_\mu$.
Elements of $C^{\infty}(M)$ are in one to one
correspondence with fields on the four dimensional space-time. The
space of one forms is defined by:
$$\mathbb{A} \ = \ f(x) [\Dir , g(x)]  \ = \ f(x) \der_{\mu}(g)(x) \gamma^{\mu}
\ \equiv \ A_{\mu}(x) \gamma^{\mu} \ ,
$$
with $A_{\mu}(x) \in C^{\infty}(M)$\footnote{Note that the functions
appearing in these relations are to be thought of as operators. Then,
for example, $\der g(x)  = \der(g)(x) + g(x) \der $}.
Note that the decomposition of one forms
this way is similar to the usual writing of one forms as
$A_{\mu}(x) \dd x^{\mu}$, only in this case the matrices $\gamma^{\mu}$
play the role of a basis in the space of one-forms. The field strength
will then be (here we use the anticommutator):
\bea{f_def}
\mathbb{F} & = & \dd \mathbb{A} \ = \ \{ \Dir,A_{\mu}(x)\gamma^{\mu} \}
\ = \ \gamma^{\nu} \gamma^{\mu} \der_{\nu} A_{\mu} +
\gamma^{\mu} \gamma^{\nu} A_{\mu} \der_{\nu} \nonumber \\
& = & \gamma^{\nu} \gamma^{\mu} \der_{\nu} (A_{\mu}) + 2 \eta^{\mu \nu}
A_{\mu} \der_{\nu} \nonumber \\
& = & \gamma^{\nu} \gamma^{\mu}{1\over 2}  [ \der_{\nu} (A_{\mu}) -
 \der_{\mu} (A_{\nu}) ] \ .
\eea In going from the second to the last line, we have
antisymmetrized the result with respect to indices $\mu, \nu$, and
dropped the symmetric terms (the ones proportional to $\eta ^{\mu \nu}$).
This prescription is
imposed by the requirement that if $\mathbb{A} = \dd \Phi$, $
\dd \mathbb{A} = 0$. In other words, the equivalence relation
necessary to get rid of the junk forms is: two forms are
equivalent to each other if their difference is symmetric
in the $\mu, \nu$ indices, and proportional to
the identity matrix. Alternately, it is equivalent to using the
wedge product of $\gamma$ matrices as
a basis in the space of higher forms; $\mathbb{F}$ can then be written
as 
$$
\mathbb{F}\ =\ {\der_{\nu}(A_{\mu})\gamma^{\nu}\wedge\gamma^{\mu}}\ ,
$$
with $\gamma^{\nu}\wedge \gamma^{\mu} = (\gamma^{\nu}\gamma^{\mu}
- \gamma^{\mu}\gamma^{\nu})/2$.

\section{Two-sheeted space-time.}

In the previous section we have given a brief description of the
general formalism of Connes to build noncommutative field
theories. There are several realizations of this formalism in
practice \cite{ConnesLott,Coquereaux,Chamseddine,Wulkenhaar}.
In this section (and the rest of the paper) we
will discuss a model, which we call two-sheeted space time 
\cite{Viet,VietWali}.

In our scenario, the spectral triple consists of:

{\bf Hilbert Space $\cal{H}$}: The  direct product of the usual
Minkowski spin manifold times $\mathbb{C}^2$ ($\mathbb{C}$ being
the algebra of complex numbers). Formally, it can be represented
as
$${\bf {\Psi}} \ = \left[ \begin{array}{c} \Psi_L \\ \Psi_R \end{array} \right], $$
where $\Psi_L,\Psi_R$ are each a Dirac spinor by itself; however,
we can choose that each such Dirac spinor has specific helicity
(either left or right-handed) and that they live on different
sheets.

{\bf Algebra $\cal{A}$}: The most general (smooth) operator acting
on such a Hilbert space is a 2$\times$2 matrix with elements
$C^{\infty}$ functions. However, we choose the algebra ${\cal A}$
to be a subset of of these matrices; specifically we require that
the representations of elements of ${\cal A}$ be diagonal
matrices:
\beq{zero_form}
 \rho(a) \ = \
  \left[ \begin{array}{cc} f_1(x) & 0 \\ 0 & f_2(x) \end{array} \right] \ .
\eeq

{\bf Dirac Operator}:  If we take this also to be diagonal, the
resulting theory would not be very interesting; the fields living
on separate sheets will not interact with each other, and in effect we
will have two copies of the standard theory on a manifold.
Therefore, we take the Dirac operator to have off-diagonal terms:
\beq{dir_op1} {\cal D} \ = \
\left[ \begin{array}{cc} i\Dir & - \gamma^5 M \\
 \gamma^5 M^\dag & i \Dir \end{array} \right] \ ,
\eeq where $M$ is a scalar operator with dimensions of mass. With
this definition, a general one-form is given by
 \beq{gen_a}
\mathbb{A} \ = \
\left[ \begin{array}{cc} \gamma^{\mu} A_{1\mu} (x) & -\gamma^5 \Phi_1 (x) \\
 \gamma^5 \Phi_2 (x) &  \gamma^{\mu} A_{2\mu}(x) \end{array} \right] \ ,
\eeq where $A_{i\mu}$ are the gauge fields, and $\Phi_i$ are the
Higgs fields of the theory.

The model described above is easily extended to  incorporate 
non-abelian gauge theories. In the general case, one can take the
fields $\Psi_L, \Psi_R$ as multiplets under a fundamental representation
of some gauge groups $G_1,G_2$. Let us denote the dimension of the
$\Psi_L$ multiplet by $n$ and the dimension of the
$\Psi_R$ multiplet by $m$. Then the $C^\infty$ functions $f_1(x),A_{1\mu}(x)$,
and $f_2(x), A_{2\mu}(x)$
will became $n\times n$ matrices, and $m\times m$, respectively.
They can be written in terms of the generators
of the corresponding Lie algebras $T_a, T'_b$:
\bea{gauge}
f_1(x) = f_1^a(x) T_a & &
A_{1\mu}(x) = A_{1\mu}^a(x)T_a  \nonumber \\
  f_2(x) = f_2^b(x) T'_b & &
A_{2\mu}(x) = A_{2\mu}^b(x) T'_b \ .
\eea

To support the identification of the $\Phi$ fields in (\ref{gen_a}) as
Higgs fields, consider the fermionic part of the Lagrangian. With the
 covariant derivative defined in the usual way,
\beq{cov_de} i \mathbb{D} \ = \ {\cal D} + \mathbb{A} \ , \eeq
one has
\bea{ferm_l}
 {\cal L}_{\psi} & = & 
  {\bf \bar{\Psi} } \ i \mathbb{D} \ {\bf {\Psi}} \nonumber \\
 & = & \bar{\psi}_L (i \Dir +\not{A_1}) \psi_L \ + \
       \bar{\psi}_R (i \Dir +\not{A_2}) \psi_R \   \nonumber \\
& & -\left(\ \bar{\psi}_L ( \Phi_1 + M) \psi_R +
 \bar{\psi}_R (\Phi_2 + M^\dag) \psi_L\  \right) \ .
\eea
In order for this lagrangian to be hermitian, we define $\Phi_2 =  \Phi_1^\dag$.

The field strength is then
\beq{fstr}
\mathbb{F} \ = \ \dd \mathbb{A} \ +\ \mathbb{A}\wedge \mathbb{A} \ ,
\eeq
where
\bea{Fdef}
 \dd \mathbb{A} & = & \{ {\cal D}, \mathbb{A} \}\nonumber \\
&  = &
\left[ \begin{array}{cc} i\Dir(\not{A_1}) - M \Phi^\dag - \Phi M^\dag & 
        \gamma^{\mu} \gamma^5 (-i\der_{\mu}(\Phi) + M A_{2\mu}- A_{1\mu} M) \\ 
        \gamma^5 \gamma^{\mu} 
                (-i\der_{\mu}(\Phi^\dag) + M^\dag A_{1\mu}- A_{2\mu} M^\dag)  &
        i\Dir(\not{A_2}) - M^\dag \Phi - \Phi^\dag M
   \end{array} \right] \ ,
\eea
and
\bea{wedp}
 \mathbb{A} \wedge \mathbb{A} & = & \frac{1}{2} \{ \mathbb{A} , \mathbb{A} \}
 \nonumber \\
 & = & \left[ \begin{array}{cc} \not{A_1} \not{A_1} - \Phi \Phi^\dag &
         \gamma^{\mu} \gamma^5 ( -A_{1\mu} \Phi + \Phi A_{2\mu} ) \\
         \gamma^5 \gamma^{\mu} (\Phi^\dag  A_{1\mu} - A_{2\mu} \Phi^\dag ) &
         \not{A_2} \not{A_2} - \Phi^\dag \Phi
 \end{array} \right] \ .
\eea
The diagonal elements of the form $\mathbb{F} $ contain $(i/2) \gamma^{\mu}
\gamma^{\nu} (F_{1,2})_{\mu\nu}$, with
$$ F_{1\mu\nu} \ = \ \der_{\mu} A_{1\nu} - \der_{\nu} A_{1\mu} - i
[A_{1\mu},A_{1\nu}] \ ,$$
and the equivalent  for $F_2$. Thus one obtains the proper covariant
expressions for the field strengths of the respective gauge fields. 
 Furthermore, it is convenient to redefine
the Higgs fields as $H = \Phi + M$; then the field strength (\ref{fstr}) 
can be written as:
\beq{Fdef2}
\mathbb{F} \ = \ 
\left[ \begin{array}{cc} {i \over 2} \gamma^{\mu} \gamma^{\nu} F_{1\mu\nu}
 + (M  M^\dag - H H^\dag) & 
        -\gamma^{\mu} \gamma^5 (i\der_{\mu}(H) - H A_{2\mu} + A_{1\mu} H) \\ 
        -\gamma^5 \gamma^{\mu} 
                (i\der_{\mu}(H^\dag) - H^\dag A_{1\mu}+ A_{2\mu} H^\dag)  &
         {i \over 2} \gamma^{\mu} \gamma^{\nu} F_{2\mu\nu}
 + ( M^\dag M - H^\dag H)
   \end{array} \right] \ .
\eeq

Some care has to be exercised here in treating the contributions of the
Higgs field to the diagonal elements. As mentioned
in the previous section, if $\del$ is a proper exterior 
derivative, $\del(\del\al) = 0$ for any form $\al$. Thus, if one takes
$\mathbb{A} = \dd \rho$, with $\rho$ an element of the algebra as in 
Eq. (\ref{zero_form}), we should have 
 $d \mathbb{A} = d d \rho = 0$. However, from Eq. (\ref{Fdef}):
$$
d d \rho \ = \ 
\left[ \begin{array}{cc} M M^\dag f_1 - f_1 M M^\dag & 0 \\ 
0 &  M^\dag M f_2 - f_2  M^\dag M \end{array} \right] \ .
$$ 
 If $M M^\dag$ is not proportional to the unity, the diagonal
elements are not generally zero. We need a prescription to deal with such
cases. A suitable prescription for our purposes, which gives rise to a 
gauge covariant Higgs potential, is to take the trace over the Higgs
field contributions
\footnote{One could also drop these contributions
completely, in effect setting the curvature associated to the Higgs fields to
zero. However, the model obtained in this case is not very interesting,
since it does not have a Higgs potential.}, that is, to
replace in Eq. (\ref{Fdef2}), 
$(M  M^\dag - H H^\dag)$  by Tr$(M  M^\dag - H H^\dag)$.
However, note that here the trace is understood to be taken only when 
it is required for the cancellation of such terms.
For example, in the case when the fermion spinor
space can be split into subspaces which do not mix
(like lepton and quark subspaces in the Standard Model), one should
take trace on each subspace separately. 

The Lagrangian for the gauge sector (including the Higgs) will then be
\beq{lg}
{\cal L}_G \ = \ -{1\over 4}\ \hbox{Tr}\ \mathbb{F} \mathbb{F} \ ,
\eeq
where the trace is first taken over the $\gamma^\mu$ matrices and 
then over the internal symmetry indices.
The $-1/4$ factor in front of the trace insures that the kinetic energy
for the gauge fields $F_1, F_2$ has the standard normalization.
Then, we find
\beq{lg_exp}
{\cal L}_G \ = \ {\cal L}_A \ + \ {\cal L}_H \ - \ V(H) \ ,
\eeq
where
\beq{la}
{\cal L}_A \ = \ -{1\over 4}\sum_{i} F_{i \mu\nu} F_i^{\mu\nu}
\eeq
with
$$ F_{1\mu\nu} \ = \ F_{1\mu\nu}^a T_a \ , \ 
F_{2\mu\nu} \ = \ F_{2\mu\nu}^b T'_b \ , $$
with normalization Tr$(T_a T_{a'})= (1/2)\delta_{aa'}$. The kinetic
energy for the Higgs fields has the form
\beq{h_ke}
{\cal L}_H \ = \ 
|\der^{\mu}(H) + i H A_{2}^{\mu} - i A_{1}^{\mu} H|^2
+ |\der_{\mu}(H^\dag) + i H^\dag A_{1\mu} - iA_{2\mu} H^\dag|^2 \ ,
\eeq
where we use the definition $|K|^2 = K^\dag K$. Finally, the Higgs
potential is given by
\beq{h_pot}
V(H)\ = \ \left(\hbox{Tr} (M  M^\dag - H H^\dag) \right)^2 \ +
 \ \left(\hbox{Tr} ( M^\dag M -  H^\dag M) \right)^2 \ ,
\eeq
which requires that at least some Higgs fields have non-zero expectation
values, thus insuring spontaneous symmetry breaking.

Note that Eq.(\ref{h_ke}) contains the proper covariant derivative for 
the Higgs fields $H$. To see this, note that the transformation properties
of the Higgs fields are defined by the requirement that the
fermionic lagrangian (\ref{ferm_l}) be gauge invariant. Thus, if the fermion
fields $\Psi_L$ form a multiplet supporting a fundamental representation of
the gauge group $G_1$, so do the fields $H$. That is, the transformation
properties of the Higgs fields under an infinitesimal gauge transformation
will be the same as for the $\Psi_L$ fermions:
$$
\Psi_L(x) \rightarrow (1 + i\alpha^a T^a) \Psi_L \ \ , \ \ 
H(x) \rightarrow (1 + i\alpha^a T^a) H \ .
$$    
Then it is easy to verify that the derivative  in Eq.(\ref{h_ke})
is the appropriate
covariant derivative, and the theory is gauge invariant.

To review what we have learned so far: we have started by taking the elements
of the algebra ${\cal A}$ to be block-diagonal matrices. However, due to the
non-diagonal structure of the Dirac operator, the one-forms associated
with the algebra ${\cal A}$ are also non-diagonal - that is, non-commutative.
The diagonal elements of a one-form $\mathbb{A}$ are gauge fields, while
the off-diagonal elements turn out to be Higgs fields. The resulting theory
turns out to be gauge invariant (one has to redefine the Higgs fields first,
though). Moreover, due to the noncommutativity of one-forms, the Higgs
sector aquires a quartic potential, thus ensuring spontaneous symmetry breaking.
The structure of the Higgs potential is determined by the structure of the
off-diagonal terms in the Dirac operator (or, since these terms are related
to the fermion masses, one can say that the Higgs potential is related
to the mass spectrum of the fermion sector).

We end this section with some comments on the gauge couplings and the 
Higgs-fermion Yukawa couplins. In the above equations we set the gauge 
couplings to be one. Different values can be introduced either directly
(that is, setting $i \mathbb{D}  =  {\cal D} + g \mathbb{A} $ in Eq. 
(\ref{cov_de}) and $\mathbb{F} = \dd \mathbb{A} + 
g \mathbb{A} \wedge \mathbb{A} $) or by generalizing the trace over 
$\mathbb{F}$ to include a gauge coupling matrix:
\beq{gauge_cpl}
{\cal L}_G \ = \ 
-{1\over 4}\ \hbox{Tr}\ 
\left[ \begin{array}{cc} 1/g_1^2 & 0 \\ 0 & 1/g_2^2 \end{array} \right]\
\mathbb{F} \mathbb{F} \ .
\eeq
Then one should use fields with proper normalization $A_1/g_1 \rightarrow A_1,
A_2/g_2 \rightarrow A_2$,  in 
effect setting the gauge couplings of the fields $A_1, A_2$ to be $g_1, g_2$
(one should also normalize the Higgs fields). Also, note that in the
case when we have only one Higgs multiplet and one fermion multiplet, the 
Yukawa coupling of the Higgs with the fermions is fixed, being
given by the Higgs field normalization constant. In the case when one 
has several fermion multiplets, one generally can (or has to) set the
Yukawa coupling for each multiplet by hand; however, 
 one still gets a sum rule relating
the sum of the Yukawa couplings squared  
to the Higgs normalization constant. If one has
several Higgs multiplets, one gets several such sum rules.

\section{Standard Model on two-sheeted space time}

As the first example of the general formalism described in the previous section,
let us discuss the implementation of the Standard Model $SU_C(3)\times
SU_L(2)\times U_Y(1)$ gauge theory. For simplicity, let us first restrict 
our analysis to a subspace of the fermion spinor space, 
particularly the one spanned by the $u,d$ quark fields. Then, the components
of the spinor multiplets would be
$$ \Psi_L \ = \ \left[ \begin{array}{c} u_L \\ d_L \end{array} \right] \ , \
\Psi_R \ = \ \left[ \begin{array}{c} u_R \\ d_R \end{array} \right] \ .
$$
As is well known, the $\Psi_L$ components form a $SU_L(2)$ doublet, while
the individual components of both $\Psi_L$ and $\Psi_R$ are $U_Y(1)$ singlets.
 
 Then, over this subspace, the gauge fields components are
\bea{sm_gf}
 A_{1\mu} & = &  A_{1\mu}^a \tau_a + {1 \over 2} Y_{QL} B_{\mu} \ , \nonumber \\
 A_{2\mu} & = & { 1 \over 2}  Y_{QR} B_{\mu} \ ,
\eea
where $A^a$ are the $SU_L(2)$ gauge fields and $B$ is the hypercharge field.
The diagonal matrices $ Y_{QL}, Y_{QR}$ contain the hypercharges of the left
and right-type quark fields:
$$
Y_{QL} \ = \ {1\over 3} 
      \left[ \begin{array}{cc} 1 & 0 \\ 0 & 1 \end{array} \right] \ , \
Y_{QR} \ = \  
      \left[ \begin{array}{cc} 4/3 & 0 \\ 0 & -2/3 \end{array} \right]\ .
$$
Note here the peculiar way we have introduced the hypercharge field. The
matrix $A_1$ now contains fields associated with two gauge symmetries. This
would generally require that $A_1$ be split into two parts: $A_1 \rightarrow
 \hbox{diag}
(A_1,B_1)$, but this is not necessary in this particular case, since the
hypercharge matrix $Y_{QL}$ is proportional to the identity matrix, and 
therefore commutes with the generators of the $SU_L(2)$
gauge transformations. This means that the $SU_L(2)$ and $U_Y(1)$ gauge
fields will 
not mix between them when computing $\mathbb{F}$ and the trace of
$\mathbb{F}\mathbb{F}$. Therefore we can add them directly in the manner of
(\ref{sm_gf}).

The Higgs matrix for the model has the form
$$
H \ = \  
 \left[ \begin{array}{cc} \bar{h}_0 & h_+ \\ -h_- & h_0 \end{array} \right]\ 
 = \ [ \widetilde{\Phi} , \Phi] \ ,
$$ 
where $\Phi$ is the $SU_L(2)$ doublet and $\widetilde{\Phi}$ is its charge
conjugate. Under an infinitesimal $U_Y(1)$ gauge transformation $\alpha(x)$
the Higgs fields will change according to
$$ H \ \rightarrow \ (1+i {\alpha \over 2} Y_{QL})\  H \ 
(1-i {\alpha \over 2} Y_{QR}) 
\ \rightarrow \ H \ (1+i {\alpha \over 2} (Y_{QL}-Y_{QR})) \ , $$
since $Y_{QL}$ is proportional to the identity matrix. 
The hypercharge matrix for the Higgs fields will then be 
$$ Y_H \ = \ Y_{QL} -   Y_{QR} \ = \ 
\left[ \begin{array}{cc} -1 & 0 \\ 0 & 1 \end{array} \right] \ ,
$$
that is, the doublet $\Phi$ and antidoublet $\widetilde{\Phi}$ will
have definite hypercharge +1 and -1 respectively. The covariant derivative
for $H$ turns out to be
$$
D_{\mu} H \ = \ \der_{\mu} H \ -\ i A_{1\mu}^a \tau_a H \ - \
i H {1\over 2 } Y_H B \ ,$$
or, in terms of the $\Phi, \widetilde{\Phi}$ doublets:
\bea{h_ke1}
D_{\mu} ( \widetilde{\Phi}, \Phi) & = & \left(
(\der_{\mu}  - i A_{1\mu}^a \tau_a   + i {1\over 2 }  B ) \widetilde{\Phi} \ ,
\right. \nonumber \\
& &
\left. 
(\der_{\mu}  - i A_{1\mu}^a \tau_a   - i {1\over 2 }  B ) {\Phi}
\right) \ .
\eea
Then, 
\bea{h_lag}
 {\cal L}_{H} & = & 2\ \hbox{Tr}\  (D^{\mu} H^\dag) (D_{\mu} H) \nonumber \\
& = & 2 (D^{\mu}  \widetilde{\Phi}^\dag) (D_{\mu}  \widetilde{\Phi}) \ + \
2 (D^{\mu} \Phi^\dag) (D_{\mu}  {\Phi})
\eea
will give the standard form for the SM Higgs kinetic energy terms, although
it still has to be normalized (the 2 factor in front comes from
the fact that there are two equal terms in Eq.(\ref{h_ke}) ).

Let us assume for a moment that this is the whole extent of the theory
(that is, there are no other fermions), 
and work out the gauge coupling constants, Yukawa couplings and normalization
factors. After multiplying with the gauge coupling matrix (as in Eq. 
(\ref{gauge_cpl})), the gauge fields have to be rescaled
\bea{ren_gf}
 & & {1 \over g_1} \ A_1^a \ \rightarrow \ A_1^a \nonumber  \\  
& & \left[ {1 \over g_1^2} \hbox{Tr} \left({Y_{QL}^2 \over 2}\right)
+ {1 \over g_2^2} \hbox{Tr} \left({Y_{QR}^2 \over 2}\right) \right]^{1/2} 
B \ \rightarrow \ B \ ,
\eea
thus setting the weak coupling constant $g = g_1$ and the
hypercharge coupling constant
$$ {1\over g'} \ = \ 
\left[ {1 \over g_1^2} \hbox{Tr} \left({Y_{QL}^2 \over 4}\right)
+ {1 \over g_2^2} \hbox{Tr} \left({Y_{QR}^2 \over 4}\right) \right]^{1/2} \ .
$$

If one wishes to give different masses to the up and down type
fermions, one has to introduce different Yukawa couplings in the Higgs matrix
$$
H \ = \  
 \left[ \begin{array}{cc} \bar{h}_0 & h_+ \\ -h_- & h_0 \end{array} \right]
\left[ \begin{array}{cc} \lam_u & 0 \\ 0 & \lam_d \end{array} \right]
\ = \ [ \lam_u \widetilde{\Phi} , \lam_d \Phi] \ ,
$$ 
since the vev of the $\widetilde{\Phi}$ doublet gives mass
to the up-type fermions, while the vev of the $\Phi$ one gives mass to
the down-type fermions. Note that this implies that the  matrix
$M$ appearing in the Dirac operator has a similar structure
$$
M \ \sim \ 
\left[ \begin{array}{cc} m \lam_u & 0 \\ 0 & m \lam_d \end{array} \right] \ .
$$

One has now to compute the rescaling factor for the Higgs field. The
kinetic energy term (\ref{h_lag}) becomes
$$
{\cal L}_{H} \ = \ \left( {1\over g_1^2} + {1\over g_2^2} \right) \lam_u^2 \
(D^{\mu}  \widetilde{\Phi}^\dag) (D_{\mu}  \widetilde{\Phi}) \ + \
\left( {1\over g_1^2} + {1\over g_2^2} \right) \lam_d^2 \
(D^{\mu} \Phi^\dag) (D_{\mu}  {\Phi}) \ ,
$$
which requires the redefinition (rescaling) of $H$
$$\left[
\left( {1\over g_1^2} + {1\over g_2^2} \right) (\lam_u^2 + \lam_d^2)
\right]^{1/2} \ H \ = \ \lam \ H 
\ \rightarrow \ H \ . 
$$
Then, the effective couplings $\lam^e_{u,d}$ to fermions will turn out to be
$$ \lam^e_u = \lam_u/\lam \ , \ \lam^e_d = \lam_d/\lam \ , $$
and we have the sum rule mentioned in the previous section
\beq{sr_sm}
(\lam^e_u)^2 +  (\lam^e_d)^2 \ = \ { g_1^2 g_2 ^2 \over g_1^2 + g_2^2} \ .
\eeq

Finally, let us consider the terms giving rise to 
 the Higgs potential. Since the matrix M is
not proportional to identity and it does not commute
with a general zero-form, we should use the trace prescription in (\ref{Fdef}).
Therefore,
\bea{hp_sm} 
V(H)\ & = & \ \left( {1\over g_1^2} + {1\over g_2^2} \right)
\left( \hbox{Tr} [M M^\dag - H H^\dag] \right)^2 \nonumber \\
& \rightarrow & \left( {1\over g_1^2} + {1\over g_2^2} \right)^{-1}
(h_0 \bar{h}_0 + h_- h_+ - m^2)^2 \ ,
\eea
in terms of the rescaled fields. From this, we see that the neutral Higgs
component field $h_0$ aquires a vacuum expectation value 
$<h_0>  = m\ ( = v/\sqrt{2})$,
and after symmetry breaking the surviving Higgs particle gets a mass
$$ m_h \ = \ 2 m \sqrt{(\lam^e_u)^2 +  (\lam^e_d)^2}$$
(where the sum-rule (\ref{sr_sm}) has been used).

We wish now to extend the previous construction to the whole 
Standard Model. One must then add leptons,
and include color and flavor. For this
purpose,  we first extend the fermion space
$$ 
\Psi_L \ = \ 
\left[ \begin{array}{c}
\left( \begin{array}{c} u_L \\ d_L \end{array} \right)_\al \\
\left( \begin{array}{c} \nu_L \\ e_L \end{array} \right)
\end{array}
\right]_i \ , \
\Psi_R \ = \ 
\left[ \begin{array}{c}
\left( \begin{array}{c} u_R \\ d_R \end{array} \right)_\al \\
\left( \begin{array}{c} 0 \\ e_R \end{array} \right)
\end{array}
\right]_i \ ,
$$
where $\al$ are the color indices, and $i$ the flavor ones.
Then the gauge one-form will split into components acting on
the lepton  and quark subspaces. On the lepton subspace, 
we  have
\bea{sm_gf_lep}
 A_{1\mu}^E & = &  \left( 
 A_{\mu}^a \tau_a + {1 \over 2} Y_{EL} B_{\mu}\right) \otimes {\bf 1}_{ij}
  \nonumber \\
  A_{2\mu}^E & = & \left( { 1 \over 2}  Y_{ER} B_{\mu} \right) 
  \otimes {\bf 1}_{ij} \nonumber \\
  H^E & = & \Phi \otimes \lam^{l}_{ij}  \ ,
\eea
where  $Y_{EL}, Y_{ER}$ matrices contain the hypercharge  numbers
associated with the left and right handed charged leptons and neutrinos,
and $\lam^{l}_{ij}$
is the Yukawa coupling matrix for leptons. On the quark subspace
\bea{sm_gf_qu}
 A_{1\mu}^Q & = &  \left( 
 A_{\mu}^a \tau_a + {1 \over 2} Y_{QL} B_{\mu}\right) 
 \otimes (G_{\mu}^b \Lambda_b)_{\al \be} \otimes {\bf 1}_{ij}
  \nonumber \\
  A_{2\mu}^Q & = & \left( { 1 \over 2}  Y_{QR} B_{\mu} \right) 
  \otimes (G_{\mu}^b \Lambda_b)_{\al \be}
  \otimes {\bf 1}_{ij} \nonumber \\
  H^Q & = &
  \widetilde{\Phi} \otimes {\bf 1}_{\al \be} \otimes \lam^{u}_{ij} \ + \
  \Phi \otimes {\bf 1}_{\al \be} \otimes \lam^{d}_{ij}  \ ,
\eea
where $G^b_\mu$ are the $SU_C(3)$ gauge bosons, and the $\Lambda_b$ are
the generators of the $SU_C(3)$ gauge transformation. Also, the 
$\lam^{u}_{ij}$ and $\lam^{d}_{ij}$ are the Yukawa coupling matrices
for the up and down-type quarks.

One can then easily verify that one obtains the Standard Model lagrangian.
The gauge couplings for the electroweak sector are given by: 
\bea{gc_fullsm}
{1\over g^2} & = & { 4 N \over g_1^2} \nonumber \\
 {1\over g'^2} & = &  
 {N \over g_1^2} \hbox{Tr} \left( { 3 Y_{QL}^2 
+ Y_{EL}^2 \over 2}\right)
+ {N \over g_2^2} \hbox{Tr} \left({3Y_{QR}^2 
+ Y_{ER}^2 \over 2}\right)  \nonumber \\
& = &  {4 N \over 3 g_1^2} + {16 N \over 3 g_2^2}
\ .
\eea
Here $N$ stands for the number of generations. The Higgs kinetic energy 
has the form
\bea{hi_ke}
{\cal L}_{H} & = & \left( {1\over g_1^2} + {1\over g_2^2} \right) 
3 \hbox{Tr} (\lam^{u})^{2} \
(D^{\mu}  \widetilde{\Phi}^\dag) (D_{\mu}  \widetilde{\Phi}) 
\nonumber \\
& & \ + \
\left( {1\over g_1^2} + {1\over g_2^2} \right) 
\left(3 \hbox{Tr} (\lam^{d})^{2}  + \hbox{Tr} (\lam^{l})^{2} \right) \
(D^{\mu} \Phi^\dag) (D_{\mu}  {\Phi}) \ .
\eea
 Therefore, the constant involving in rescaling the Higgs field is 
$$ \lam^2 \ = \
\left( {1\over g_1^2} + {1\over g_2^2} \right)
\left(3 \hbox{Tr} (\lam^{u})^{2} + 
3 \hbox{Tr} (\lam^{d})^{2}  + \hbox{Tr} (\lam^{l})^{2} \right)
\ \simeq \ 3 \lam_t^2 \left( {1\over g_1^2} + {1\over g_2^2} \right) \ ,
$$ 
if we assume that the top Yukawa coupling dominates. The resulting sum rule
will predict that the effective top Yukawa coupling is
$$ \lam_t^{e2}  \ = \ {1\over 3}{g_1^2 g_2^2 \over g_1^2 + g_2^2}
\ = \ {16 N \over 9}{g^2 g'^2 \over g^2 + g'^2} \ ,
$$ 
which, if one uses the values for the electroweak couplings at 
$M_Z$ scale $\al_1(M_Z) = 1/58.97,\ \al_2(M_Z) = 1/29.61$ comes out to
$\lam_t^e \simeq 0.8$ (for the number of generations $N=3$). 
This is to be compared with
the Standard Model value, $\lam_t^{SM} \simeq 1$.

Finally, the Higgs potential is
\bea{hipot_sm}
 V(H) & = & \left( {1\over g_1^2} + {1\over g_2^2} \right)
\left[ (h_0 \bar{h}_0 + h_+ h_ - -\lam^2 m^2) 
\left(3 \hbox{Tr} (\lam^{u})^{2} + 
3 \hbox{Tr} (\lam^{d})^{2}  + \hbox{Tr} (\lam^{l})^{2} \right) \right]^2
\nonumber \\
& \rightarrow & \left( {1\over g_1^2} + {1\over g_2^2} \right)^{-1}
(h_0 \bar{h}_0 + h_+ h_ - - m^2)^2 \ .
\eea
Then one would obtain a Higgs mass:
$$ M_h \ \simeq  \ 2 \sqrt{3} \  \lam_t^e m \ \simeq \ 3.5 \ m_t \ .$$

One notes that this model predicts a rather large Higgs mass, Also, 
it is not exactly in agreement with the Standard Model value for
the top mass (although it is not too far either)\footnote{It is
possible to fix the prediction for the top mass by introducing
a separate coupling for the Higgs part of the field strength (as, for
example, in \cite{Viet}). However, this would lead to a loss of 
predictive power for the model.}. However, this is not too troubling;
we do not after all expect that the Standard Model is a fundamental theory,
valid at all scales. Instead, it seems probable that the 
$SU_C(3)\times SU_L(2) \times U_Y(1)$ gauge symmetry visible at $M_Z$ scale
is a remnant of the breaking of some larger symmetry, which
is manifest at a higher scale. Hence, one should consider  applying
the framework used above to such extensions of the Standard Model. 

\section{Left-Right symmetric model on two-sheeted space time}

In spite of the great phenomenological success of the standard model
as a gauge theory based on spontaneously broken symmetry
$SU_C(3)\times SU_L(2) \times U_Y(1)$,  
it is recognized that it has several unsatisfactory features that include,
 besides the proliferation of free parameters, a lack of an understanding of
the origin of parity violation in the low energy region.
  Left-right symmetric models have always been
attractive as the minimal extensions
of the standard model. 
A natural consequence of left-right symmetric models is the 
existence of right-handed neutrinos, which, through the 
seesaw mechanism \cite{seesaw},
give rise to small mass for the left handed neutrinos. 
The discovery of convincing evidence for
non-zero neutrino masses gives therefore further weight to the
ideea that the Standard Model may be a low energy version of such models.



To start formulating the left-right symmetric model as a theory on
two-sheeted space time, we have to first specify the Hilbert space of
spinors. One could use the same space as for the Standard Model;
however, in that case the only Higgs multiplets one can introduce
are $SU_L(2)\times SU_R(2)$
bidoublets. Since one needs triplet Higgses to break the left-right
symmetry (as well as to give mass to the
right handed neutrinos), we take 
the Hilbert space to also include
the charge conjugate fields of $\Psi_L, \Psi_R$,
$$ 
\Psi\ = \ 
\left[ \begin{array}{c}
\left( \begin{array}{c} \Psi_L \\ \Psi_R^c \end{array} \right) \\
\left( \begin{array}{c} \Psi_R \\ \Psi_L^c \end{array} \right)
\end{array}
\right]_i \ ,
$$
in such a way that the left-handed fields live on one sheet and 
the right-handed ones live on the other. The matrix structure associated
with the gauge fields will then be
\bea{a_comp}
 A_{1\mu}  & = & 
\left[ \begin{array}{cc} A_{L\mu}^a \tau_a + {Y_L } B_\mu/2 & 0 \\
 0 & A_{R\mu}^a \tau_a - {Y_R } B_\mu /2\end{array} \right]
\nonumber \\
A_{2\mu}  & = & 
\left[ \begin{array}{cc} A_{R\mu}^a \tau_a + {Y_R } B_\mu/2 & 0 \\
 0 & A_{L\mu}^a \tau_a - {Y_L } B_\mu/2 \end{array} \right] \ .
\eea
The $A_L^a, \ A_R^a$ are the gauge fields associated with the $SU_L(2), \
SU_R(2)$ group transformations, while $B$ is the one associated with
the $U(1)_{B-L}$ group. The $Y_{L,R}$ numbers are the charges associated
with $U(1)_{B-L}$ transformations for the left and right-handed fields. They
take different values on lepton and quark spaces; for leptons
$Y_{EL} = Y_{ER} = -1$, and  for quarks $Y_{QL} = Y_{QR} = 1/3$.

Since charge conjugate fermions are part of the spinor space,
one can
introduce triplet Higgses in the gauge one-form. Thus, on the
lepton subspace we define
\beq{higgs_lr} \mathbb{H} \ = \  
\left[ \begin{array}{cc} H_1 \otimes \lam^{1l}_{ij} & \Delta_L 
\otimes \lam^{L}_{ij} \\
 \Delta_R^\dag \otimes \lam^{R\dag}_{ij} & H_2^c \otimes \lam^{2l}_{ij} 
 \end{array} \right] \ ,
\eeq
where $H_1$ is a (2,2,0) bidoublet and $H_2$ its conjugate,
$$ H_1 \ = \ \left[ \begin{array}{cc} h_1^0 & h_2^+ \\
 h_1^- & h_2^0 \end{array} \right] \ , \ \ 
 H_2 \ = \ \left[ \begin{array}{cc} h_2^{0*} & -h_1^+ \\
 -h_2^- & h_1^{0*} \end{array} \right] \ , \
 \ H_2^c = \sig_2 H_2^T \sig_2 \ ,
$$
and $\Delta_{L,R}$ are $SU(2)$ triplets which couple to the left
and right-handed leptons
$$
\Delta_{L,R} \ = \ \left[ \begin{array}{cc} \del^- /\sqrt{2} & \del^0  \\
 \del^{--} & -\del^- /\sqrt{2} \end{array} \right]_{L.R} \ .
$$
The $ \lam^{1,2l}_{ij},  \lam^{R,L}_{ij}$  
matrices acting on the flavor space are the Yukawa couplings associated
with  the Dirac mass term for both the charged leptons and neutrinos (the 
$\lam^{1,2l}$ matrices), and the Majorana mass terms for the right- and
left-handed neutrinos (the $\lam^{R,L}$ ones)\footnote{Note that 
in order to have left-right symmetry in the fermion sector, one
should also require that $\lam^R = \lam^L, \lam^l = \lam^{l\dag}$.}. 
On the other hand
the Higgs matrix (\ref{higgs_lr})
on the quark subspace will contain only the diagonal $H_1, H_2^c$ fields
(with Yukawa matrices $\lam^{1,2q}$ associated with the quarks) since the
$\Del_L, \Del_R$ triplets do not couple to quarks. Note also that
$H_2^c = H_1^\dag$; we use this notation to remind ourselves that the
mass term $\bar{\psi}_R^c H_2^c \psi_L^c $ is usually written
as $\bar{\psi}_L H_2 \psi_R $.

Substituting (\ref{a_comp}) and (\ref{higgs_lr}) in Eq. (\ref{gen_a})
(with $\Phi_1 = \mathbb{H}, \Phi_2 = \mathbb{H}^\dag$), one
obtains the corresponding expressions for the field strength $\mathbb{F}$,
which will give the standard Lagrangian terms for the 
gauge fields $A_L, A_R$, and $ B$.
It can easily be verified that Eq. (\ref{h_ke}) will give the right kinetic
energy terms for the Higgs  multiplets
\bea{h_ke_lr}
{\cal L}_H &  = & |\der_\mu H + i H A^a_{R\mu} \tau_a  - 
i A^a_{L\mu} \tau_a H |^2 
 \nonumber \\
{\cal L}_\Delta &  = &
|\der_\mu \Delta_{L,R} + i \Delta_{L,R} A^a_{L,R\mu} \tau_a  - 
i A^a_{L,R\mu} \tau_a \Delta_{L,R} - i  Y_E B_\mu \Delta_{L,R}|^2 \ ,
\eea
where, due to left-right symmetry, $Y_{EL} = Y_{ER} = Y_E$.
The  coupling constants for the $SU(2)$ and $U(1)$ gauge fields are given by
\bea{gc_lr}
{1\over g^2} & = & 4 N \left({ 1 \over g_2^2} + { 1 \over g_1^2}\right)
         \nonumber \\
 {1\over g'^2} & = &  
 2N \left({ 1 \over g_1^2} + { 1 \over g_2^2}\right)
   \left( { 3 Y_{Q}^2 
+ Y_{L}^2 }\right)
\ .
\eea
It is interesting to note that, due to left-right symmetry on each
sheet, the ratio of the two coupling constants  is fixed: 
$g^2/g'^2 = 2/3$ (of course, this is taken to be valid at the energy
where this description holds).  This symmetry could be broken by considering
more complex structure for the gauge coupling matrix in (\ref{gauge_cpl}).
Also, note that the rescaling constants for the Higgs fields are
\bea{higgs_rs}
\lam_{\Del R}^2 & = & \hbox{Tr} \left[ \lam^R \lam^{R\dag} \right] /(4 N g^2)
         \nonumber \\
\lam_{\Del L}^2 & = & \hbox{Tr} \left[ \lam^L \lam^{L\dag} \right] /(4 N g^2)
         \nonumber \\
 \lam_{H}^2 & = &   \hbox{Tr} \left[ \lam^{1l} \lam^{1l\dag}
 +  \lam^{2l} \lam^{2l\dag}  + 3( \lam^{1q} \lam^{1q\dag}
 +  \lam^{2q} \lam^{2q\dag}) \right]/(4 N g^2)
\ .
\eea

Before continuing, let us briefly review the salient features of
gauge symmetry breaking in the left-right symmetric models. The theory
has five gauge bosons, two charged ones ($A_L$ and $A_R$) and three
neutral: $B, A_{0L}, A_{0R}$. At some high energy, the symmetry
is unbroken, and all gauge bosons are massless. However, at some large
scale, the right-handed triplet Higgs aquires a vacuum expectation value
$<\Delta_R> = v_R$, and breaks left-right symmetry. One of the charged
bosons (more precisely, a combination of the $A_L$ and $A_R$) and one
of the neutral ones aquire masses of order $v_R$.  Since these massive
bosons have not been observed at present day colliders, one infers that
$v_R$ should be above the TeV scale\footnote{Constraints 
on the scale of $v_R$ can
 also be inferred from weak interaction data; see for
example \cite{Mohapatra} and references therein.}. 
Furthermore, the right handed neutrinos also 
aquire Majorana mass through the vev of $\Del_R$.
If one accepts that the smallness of left handed
neutrino masses is due to the seesaw mechanism, this would point
to the scale $v_R$ to be around $10^{14}$ GeV. The remaining symmetry
(that of the Standard Model $SU_L(2)\times U_Y(1)$) is broken
at the electroweak scale by vacuum expectations for the bidoublet Higgses
$<h_1^0> = v_1, <h_2^0> = v_2$. Note that the left-handed triplet
$\Del_L$ may also aquire a vacuum expectation value $v_L$; however, since
this vev gives Majorana mass to the left-handed neutrinos, $v_L$ should be
below eV scale. 
 
A very interesting question then arises: is this pattern of symmetry 
breaking consistent with our formulation of the left-right symmetric
model on a two-sheeted space time? Note that, in our scenario, the
Higgs potential is fixed  (up to maybe an overall multiplicative constant)
once the  matrix $M$ appearing in the Dirac operator has been given.
Therefore one can potentially hope to predict how the gauge symmetry is broken.
We shall investigate this in what follows, 
under different sets of assumptions concerning the structure of the $M$
matrix.

{\bf A)}
We first choose a form for the  matrix $M$ suggested by the
pattern of fermion masses (that is, choose a nonzero entry in the mass matrix
at places where fermion masses would naturally appear). Thus, one can choose
\beq{mass_patt}
M\ = \ 
\left[ \begin{array}{cc}
\left( \begin{array}{cc} m_1 & 0 \\ 0 & m_2 \end{array} \right) 
			\otimes \lam^{1l}_{ij} \ , &
\left( \begin{array}{cc} 0 & m_L \\ 0 & 0  \end{array} \right) 
			\otimes \lam^{L}_{ij} \\
\left( \begin{array}{cc} 0 & 0 \\ m_R & 0 \end{array} \right) 
			\otimes \lam^{R}_{ij} \ , &
\left( \begin{array}{cc} m_1 & 0 \\ 0 & m_2 \end{array} \right) 
			\otimes \lam^{2l}_{ij}
\end{array}
\right] \ ,
\eeq
on the lepton subspace. On the quark subspace, we will have a similar 
structure (with the fermion Yukawa couplings $\lam^{1l}, \lam^{2l}$
replaced by the quark ones $\lam^{1q}, \lam^{2q}$),
with the difference that the off diagonal matrices will be zero
(since one generally does not introduce triplets in the quark sector, and
thus one does not get a Majorana mass for the quarks).

Before computing the Higgs potential, one has to insure the $\dd \dd \rho = 0$
condition by using the trace prescription 
discussed in section I. Following Eq. (\ref{zero_form}),
if we set
$f_1 = \hbox{diag}\{f_{1L}, f_{1R}\}$, we obtain on the left-handed sheet
\beq{ddro} (\dd \dd \rho)_{11} \ = \
\left( \begin{array}{cc} 
(MM^\dag)_{11} f_{1L} - f_{1L} (MM^\dag)_{11} \ \ & 
 (MM^\dag)_{12} f_{1R} - f_{1L} (MM^\dag)_{12} \\ 
 (MM^\dag)_{21} f_{1L} - f_{1R} (MM^\dag)_{22} \ \ &
 (MM^\dag)_{22} f_{1R} - f_{1R} (MM^\dag)_{22} 
   \end{array} \right) \ ,
\eeq
with a similar form on the right-handed sheet.
In order to satisfy $\dd \dd \rho = 0$, one then should drop the off-diagonal
terms and take the trace of the diagonal ones. The contribution of the
Higgs terms to the diagonal elements of $\mathbb{F}$ is then
\bea{Higgs_2f}
&  MM^\dag - \mathbb{H}\mathbb{H}^\dag \ \rightarrow \ &
\nonumber \\
& \left[ \begin{array}{cc} 
\hbox{Tr}[H_1 H_1^\dag +  \Delta_L \Delta_L^\dag -(MM^\dag)_{11}] & 0 \\
 0 & 
 \hbox{Tr}[H_2^c H_2^{c\dag} +  \Delta_R^\dag \Delta_R -(MM^\dag)_{22}]
   \end{array} \right] & \ .
\eea 

Now, if we assume that by suitable transformations, one can take the
vacuum expectation values of the Higgs fields to be
$$
<H_1> \ = \ \left[ \begin{array}{cc} v_1 & 0 \\
0 & v_2 \end{array} \right]  , \ \
<\Delta_{L,R}> \ = \ \left[ \begin{array}{cc} 0 & v_{L,R} \\
0 & 0 \end{array} \right] \ ,
$$
the Higgs potential will become
\bea{higgs_vev1}
V(<\mathbb{H}>) & = & \left[ (v_1^2 + v_2^2 - m_1^2- m_2^2 )|\lam^{1l}|^2 
           + (v_L^2 - m_L^2)|\lam^{L}|^2 \right]^2 \nonumber \\
         & &  + \left[ (v_1^2 + v_2^2 - m_1^2- m_2^2 )|\lam^{2l}|^2 
           + (v_R^2 - m_R^2)|\lam^{L}|^2 \right]^2
         \nonumber \\
        & & + 3 \left(v_1^2 + v_2^2 - m_1^2- m_2^2 \right)^2
        \left(|\lam^{1q}|^4 + |\lam^{2q}|^4\right) \ ,
\eea
where the first two lines are contributions coming from the lepton subspace
and the last line is the contribution due to the quark subspace. (Here we have
used $|\lam|^2$ as a shorthand for Tr$[\lam \lam^\dag]$.) One then sees that
this potential has a minimum at $v_L = m_L, \ v_R = m_R $ and
$v_1^2 + v_2^2 = m_1^2 + m_2^2$. In other words,  the vevs of the
triplets $\Delta_R, \Del_L$ can be fixed to $m_R, m_L$, but there is a
degeneracy in determining the individual values of the bidoublet Higgs
vevs. If, however, one takes $m_1^2 + m_2^2$ to be at electroweak scale,
both $v_1$ and $v_2$ are at this scale, and this does not create any problem
with respect to the symmetry breaking pattern.
So we see that we can obtain the desired symmetry breaking pattern, although
one has to put in by hand a large value for $m_R$, very small value
for $m_L$ and electroweak scale for $m_1, m_2$. 

Note that this result is by no means trivial. For example, if there would
be no quark sector in the theory (drop the third line in 
Eq. (\ref{higgs_vev1})),
one could not separate the vevs of the bidoublets from the vevs of the
triplets. Hence there would be no way to make sure that $H$ does not aquire
a $v_R$ scale vev. This is due to the highly degenerate structure of the
potential (there are several flat directions).  In particular, note
that if one adopts the pattern  (\ref{mass_patt}) for the mass matrix
$M$ (where nonzero entries correspond to Higgs fields that one expects
to get vacuum expectation values), it would be tempting to say that
$<\mathbb{H}> = M$ will give a minimum of the potential. One 
could then say that the fields $\Phi$ which appear initially in the gauge
one-form (\ref{gen_a}) are the Higgs fields in the broken symmetry
phase of the theory \cite{Chamseddine}. However, while it is true
that $V(<\mathbb{H}> = M) = 0$, this does not make it a true vacuum, since
degeneracies could exist.   

{\bf B)}
One might then also consider structures for the $M$ matrix which are not linked
to the fermion masses. Actually, the parameters which seem to have physical
relevance are the matrices  $M M^\dag$ and $M^\dag M$, since they appear in
both,  the Higgs potential and the evaluation of the $\dd \dd \rho =0$
condition.
 In the most general case, the Higgs potential can be written therefore in
terms of four parameters: $\mu_1^{l2} = \hbox{Tr} (M M^\dag)_{11},\
\mu_2^{l2} = \hbox{Tr} (M M^\dag)_{22}$ on the lepton space, and
$\mu_1^{q2} = \hbox{Tr} (M M^\dag)_{11},\
\mu_2^{q2} = \hbox{Tr} (M M^\dag)_{22}$ on the quark space. One would then
obtain
\bea{higgs_vev2}
V(<\mathbb{H}>) & = & \left[ (v_1^2 + v_2^2  )|\lam^{1l}_e|^2 
           + v_L^2  - \mu_1^{l2} \right]^2 \nonumber \\
        & & + \left[ (v_1^2 + v_2^2 )|\lam^{2l}_e|^2 
           + v_R^2 - \mu_2^{l2} \right]^2 \nonumber \\
        & & + 3 \left[(v_1^2 + v_2^2 ) |\lam^{1q}_e|^2  -\mu_1^{q2} \right]^2
        + 3 \left[(v_1^2 + v_2^2 ) |\lam^{2q}_e|^2  -\mu_2^{q2} \right]^2 \ ,
\eea
where we have used the vevs rescaled by the corresponding Higgs fields
factors (\ref{higgs_rs}), $v_{1,2} \rightarrow v_{1,2}/\lam_H$,
$v_{L,R} \rightarrow v_{L,R}/\lam_{\Del L,R}$, 
and the constants $\lam_e^{kl,q} = \lam^{kl,q} /\lam_H$. Again we see that
the scale of $v_1,\ v_2$ is set by the $ \mu^q_1, \mu^q_2$ parameters, so
these have to be at electroweak scale. Interestingly, for this potential
one does not have to necessarily
fine tune the vev of the $\Del_L$ field anymore; it will be naturally
driven  to zero for a whole range of values for the parameter $\mu_1^l$.
Indeed, let us call $v_0^2$ the value of $v_1^2 + v_2^2$ for which the last
line of Eq. (\ref{higgs_vev2}) is minimised; one then sees that if
$ v_0^2 |\lam^{1l}_e|^2 > \mu_1^{l2} $, then the minimum of the potential
requires $v_L = 0$.

{\bf C)}
In the two examples shown above, the Higgs potential has a relatively simple 
structure (being the sum of several perfect squares). More complex potentials
can be obtained in the case when the matrix $M$ has some symmetries. For 
example, if $M$ is proportional to identity in the generation space, one
need not take trace in (\ref{higgs_vev1}) over the generation indices. Then,
for example, if one would compute the contribution of the $\mathbb{H}_{11}$
part on the lepton subspace, one would obtain
$$
V_{11}(v_1,v_2,v_L) \ =  \ \hbox{Tr}
\left[ \left|\ (v_1^2 + v_2^2  )\lam^{1l}_e \lam^{1l\dag }_e 
           + v_L^2 \lam^{L} \lam^{L\dag } - \mu_1^{l2} \ \right|^2 \right] \ .
$$
One sees that while the coefficients of the potential terms 
quadratic in the Higgs
fields are proportional to Tr$(\lam \lam^\dag)$, the coefficients
of the quartic terms are proportional to Tr$(\lam \lam^\dag \lam \lam^\dag)$,
and therefore somewhat independent. One can even break the symmetry between
the vacuum expectation values of the $\phi_1^0$ and $\phi_2^0$ fields; if 
the $M_{11}$ and/or $M_{22}$ elements are taken to be diagonal in SU(2) space,
then the diagonal elements of (\ref{ddro}) will be zero, and no trace is 
necessary. In this case,  the (partial)
Higgs potential will be
$$
V_{11}(H) \ =  \ \hbox{Tr}
\left[ \left|\ H_1 H_1^\dag \otimes \lam^{1l}_e \lam^{1l\dag }_e
+  \Delta_L \Delta_L^\dag \otimes \lam^{L} \lam^{L\dag } - 
\mu_1^{l2} {\bf 1}\otimes {\bf 1} \ \right|^2 \right] \ ,
$$
or
$$
V_{11}(v_1,v_2,v_L) \ =  \ \hbox{Tr}
\left[ \left| v_1^2 \lam^{1l}_e \lam^{1l\dag }_e 
           + v_L^2 \lam^{L} \lam^{L\dag } - \mu_1^{l2}  \right|^2  
  +   \left| v_2^2 \lam^{1l}_e \lam^{1l\dag }_e - \mu_1^{l2}  \right|^2
           \right] \ .
$$
Finally, if one sets the off-diagonal elements of the $M$ matrix to zero
($M_{12} = M_{21}= 0$),
one can keep the off-diagonal elements in 
$\mathbb{H} \mathbb{H}^\dag - M M^\dag$, which will contribute to the 
Higgs potential a term
\beq{v_offd}
V_{12}(H) \ =  \ 2 \hbox{Tr}
\left[ \left|\ H_1 \Del_R \otimes \lam^{1l}_e \lam^{R}
+  \Del_L  H_1 \otimes  \lam^{L} \lam^{2l\dag}_e
\ \right|^2 \right] \ .
\eeq

It is instructive to compare the Higgs potential obtained in our model
with the general Higgs potential discussed in \cite{seesaw,Mohapatra}. We see that
our model can give rise to most of the terms found in  \cite{seesaw}.
One might note that only $H_1, H_1^\dag$ appears in our potential;
however, this is due to the particular choice of the Higgs matrix
(\ref{higgs_lr}); a more general choice

\beq{higgs_lrg} \mathbb{H} \ = \  
\left[ \begin{array}{cc} H_1 \otimes \lam^{1l}  + H_2 \otimes \lam^{3l}
& \Delta_L 
\otimes \lam^{L} \\
 \Delta_R^\dag \otimes \lam^{R\dag} & H_2^c \otimes \lam^{2l} +
 H_1^c \otimes \lam^{4l}
 \end{array} \right] \ 
\eeq
can be made, which will introduce $H_2, H_2^\dag$ in the potential 
(if desirable). The only important difference seems to be that the
terms coupling the vacuum expectation values of the left
and right triplets 
$\hbox{Tr} (\Del_L^\dag \Del_L) \hbox{Tr} (\Del_R^\dag \Del_R)$ are 
missing in our model. In the absence of such a term, one cannot take the
Higgs potential left-right symmetric ($\mu_1^{l2} = \mu_2^{l2},
\lam^L = \lam^R $). Indeed,
it turns out that for a left-right symmetric potential (as  in 
in \cite{seesaw}) what drives $v_L$ to zero while keeping $v_R$ at GUT scale
is just such a term $\sim v_L^2 v_R^2$, that couples the two vacuum
expectation values. Lacking such a term in our model, we are forced to 
take $\mu_2^l$ at GUT scale, and $\mu_1^l$ at electroweak scale.

\section{Conclusions}

We have developed a formalism based on the framework of Connes'
non-commutative geometry (NCG),
with the purpose of studying the standard model of
electroweak interactions and beyond.
Our model is based on a two-sheeted space-time that can be thought of as a
discretized version of Kaluza-Klein theory, in which the compact fifth
dimension is replaced by two discrete points. The left- and right- chiral
spinor fields live on the two separate sheets, while the gauge and Higgs
fields are part of a generalized gauge operator represented by a 
2$\times$2 matrix acting on the internal (discrete) space.

The main virtue of this framework
is that the scalar Higgs fields are an integral part of the
gauge sector.  Their gauge invariant kinetic parts in the Lagrangian as
well as quartic forms of Higgs potentials arise naturally.  Furthermore,
the possibility of spontaneous symmetry breaking is built naturally
into the Higgs potential.


While different NCG formulations of the Standard Model
 (and other gauge theories)
have been studied extensively by several authors, we
have taken in this paper a less mathematical approach
and focused more on the physics of the model.  The formalism 
allows for an easy (and transparent) construction of the Higgs sector.
The choice of Higgs multiplets appearing in the theory is dictated
by the choice of the underlying spinor space.
The Higgs potential can be written
in terms of the Yukawa couplings of the fermions and the
elements  of the matrix $M$ appearing in the
definition of the generalized Dirac operator 


Besides the predictive power in the Higgs sector,
 the NCG formalism leads to  sum rules for the Yukawa coupling constants 
of the fermions.
In the case of the minimal standard model, this will lead to a prediction
 of the top quark mass (as well as the Higgs mass). 
However, such results at the tree level form of the
Lagrangian cannot be taken seriously unless one knows at which scale
this picture holds. If NCG-inspired  theories are a description
of reality, one might expect them to be valid at scales close to
$M_{Pl}$. However, at such high scales, there are good reasons to believe that
the gauge group is a larger one, corresponding to a grand unified theory of
which the Standard Model is just the low energy limit. 

 
We are led therefore to consider higher symmetries. As an example,
in this paper we analyze the implications of two-sheeted space time picture
for the  left-right symmetric model.
Our approach predicts  several specific
forms possible for the Higgs potential. 
Interestingly, these predictions
allow the desired left-right symmetry breaking
pattern leading to the standard model. We discuss the various scenarios for
this potential depending upon the choice and the symmetries in the $M$
matrix of the Dirac operator. A more detailed quantitative study of these
features is desirable to draw concrete conclusions. But the model has
several attractive features, and fewer parameters compared with the
standard left-right symmetric models.

\section*{Acknowledgments}
We thank Nguyen Ai Viet and Alexandr Pinzul for many helpful discussions.
This work is supported  in part by the US Department of Energy 
grant number DE-FG02-85ER40231.

\end{document}